\documentclass[11pt,a4paper]{article}
\usepackage{jcappub}

\usepackage{aas_macros}
\usepackage{graphicx}
\usepackage{dcolumn}
\usepackage{bm}
\usepackage{amsmath,amssymb}
\usepackage{hyperref}

\title{Fuzzy dark matter soliton core hosting a supermassive black hole as a dense low-mass perturber in strong gravitational lensing}
\author[a,b]{Masamune Oguri}
\author[b,c]{and Naoi Kubo}
\affiliation[a]{Center for Frontier Science, Chiba University, 
  1-33 Yayoicho, Inage, Chiba 263-8522, Japan}
\affiliation[b]{Department of Physics, Graduate School of Science,
  Chiba University, 
  1-33 Yayoicho, Inage, Chiba 263-8522, Japan}
\affiliation[c]{Institute for Solid State Physics, The University of
  Tokyo, Kashiwa, Chiba 277-8581, Japan} 
\emailAdd{masamune.oguri@chiba-u.jp}
\abstract{Recent high-resolution imaging observations of strong lens
  systems reveal dense low-mass perturbers. We propose a soliton core,
  whose central density is boosted by a supermassive black hole
  (SMBH), in the fuzzy dark matter (FDM) model as an efficient
  perturber in strong gravitational lensing. The higher central
  density makes it less efficient in the tidal mass loss, and leads to
  the higher impact in gravitational lensing. We show that the mass
  profile of a $\sim 10^6M_\odot$ perturber in JVAS B1938+666, which
  does not resemble any known astronomical object, can be well
  explained by a soliton core in the FDM model with the mass of
  $4\times 10^{-21}$eV hosting an SMBH with the mass
  of $4\times 10^5M_\odot$. The high mass of the SMBH may be explained
  by several scenarios that predict heavy SMBH seeds such as the
  direct collapse black hole formation and primordial black holes.
}
\begin{document}
\maketitle

\section{Introduction}
The observation of the distribution of dark matter at the small scale
provides an important clue to the nature of dark matter
\cite{2017ARA&A..55..343B}. Strong gravitational lensing serves as one
of the most powerful methods to constrain the dark matter distribution
at the very small scale. Recently, by making use of the so-called
gravitational imaging technique \cite{2005MNRAS.363.1136K},
Refs.~\cite{2025NatAs...9.1714P,2025MNRAS.544L..24M} claim the
significant detection of a dark object, which perturbs the observed
shape of the lensed arc, with the mass of $\sim 10^6M_\odot$ toward
the strong lens system JVAS B1938+666, with the significance of $\sim
26\sigma$. The mass profile of the perturber is 
studied in detail in Ref.~\cite{2026NatAs..10..440V}, which concludes that the mass
profile of the perturber is best fitted by a two component model
consisting of a point-mass object and an extended object with a nearly
constant surface density out to a truncation radius of $\sim 140$~pc.
It is argued that the mass profile does not resemble that of any known
astronomical object, and may be explained by self-interacting dark
matter with the large effective cross-section of $\gtrsim
800~\mathrm{cm}^2\mathrm{g}^{-1}$ \cite{2026NatAs..10..440V}.
The detection of a similarly compact perturber in another strong lens
system \cite{2021MNRAS.507.1662M,2025MNRAS.540..247E} may suggest that
such dark compact perturbers are ubiquitous.  

In this paper, we argue that the mass profile of the $\sim 10^6M_\odot$
perturber in JVAS B1938+666 can be naturally explained in the fuzzy
dark matter (FDM) scenario 
\cite{2000PhRvL..85.1158H,2014NatPh..10..496S,2014PhRvL.113z1302S}. In
this scenario, the point mass and the extended components are
explained by a supermassive black hole (SMBH) and a soliton core,
respectively. We discuss parameters that reproduce the observed mass
profile. While Ref.~\cite{2025ApJ...991L..27L} also discusses a soliton
core as a possible origin of dense perturbers in strong lens systems, our
analysis differs from Ref.~\cite{2025ApJ...991L..27L} in that we 
consider the effect the SMBH that can significantly affect the mass
profile of the soliton core. We also note that,
while Ref.~\cite{2025ApJ...991L..27L} analyze the same strong lens
system JVAS B1938+666, they focus on the higher mass perturber with
the mass of $\sim 10^9M_\odot$.

This paper is organized as follows. In Sec.~\ref{sec:soliton}, we
summarize basic properties of soliton cores and discuss parameters
that explain the $\sim 10^6M_\odot$ perturber in JVAS B1938+666.
We discuss possible scenarios for the origin of the $\sim 10^6M_\odot$
perturber in Sec.~\ref{sec:scenarios}. Finally we give a summary in
Sec.~\ref{sec:summary}. Throughout the paper we assume a flat Universe
with the matter density parameter $\Omega_{\mathrm{m}}=0.3156$,
the baryon density parameter $\Omega_{\mathrm{b}}=0.05$, the
dimensionless Hubble constant $h=0.6727$, the spectral index
$n_{\mathrm{s}}=0.96$, and the normalization of the matter power
spectrum $\sigma_8=0.81$.

\section{Properties of soliton cores}\label{sec:soliton}

\subsection{Generic properties}
The governing equation of the structure
formation in the FDM model is the Schr\"{o}dinger-Poisson
equation. The soliton core, which is seen in cosmological FDM
simulations, corresponds to the ground state solution of the
Schr\"{o}dinger-Poisson equation. A commonly used fitting form of the
density profile of the soliton core is 
\cite{2014NatPh..10..496S,2014PhRvL.113z1302S} 
\begin{equation}
  \rho_{\mathrm{sol}}(r)=\frac{\rho_{\mathrm{c}}}{\left\{1+0.091(r/r_{\mathrm{c}})^2\right\}^8},
\label{eq:rho_sol}
\end{equation}
where $\rho_{\mathrm{c}}$ is given by
\begin{equation}
  \rho_{\mathrm{c}}=0.019a^{-1}\left(\frac{mc^2}{10^{-22}\,\mathrm{eV}}\right)^{-2}\left(\frac{r_{\mathrm{c}}}{\mathrm{kpc}}\right)^{-4}M_\odot\mathrm{pc}^{-3},
\label{eq:rho_c}
\end{equation}
with $a$ being the scale factor and $m$ being the mass of the
FDM particle. The core radius $r_{\mathrm{c}}$ is known to scale with
the halo mass $M_{\mathrm{h}}$ as \cite{2014PhRvL.113z1302S}
\begin{gather}
  r_{\mathrm{c}} = 1.6 g(z)\left(\frac{mc^2}{10^{-22}\,\mathrm{eV}}\right)^{-1}
  \left(\frac{M_{\mathrm{h}}}{10^9M_\odot}\right)^{-1/3}\mathrm{kpc},\label{eq:rc}\\
  g(z)=a^{1/2}\left(\frac{\zeta(z)}{\zeta(0)}\right)^{-1/6},
\end{gather}
and $\zeta(z)$ is the virial overdensity. At the lens redshift of
JVAS B1938+666, its value is $g(0.881)\simeq 0.79$. The total mass of
the soliton core is computed as 
\begin{equation}
  M_{\mathrm{sol}} = \int_0^\infty \rho_{\mathrm{sol}}(r) 4\pi r^2 dr
  \simeq 2.2\times 10^8a^{-1}\left(\frac{mc^2}{10^{-22}\,\mathrm{eV}}\right)^{-2}\left(\frac{r_{\mathrm{c}}}{\mathrm{kpc}}\right)^{-1}M_\odot.
\end{equation}
By plugging Eq.~\eqref{eq:rc} into the above equation and adopting
the lens redshift $z=0.881$, we obtain
\begin{equation}
M_{\mathrm{sol}} = 3.3\times 10^8\left(\frac{mc^2}{10^{-22}\,\mathrm{eV}}\right)^{-1}\left(\frac{M_{\mathrm{h}}}{10^9M_\odot}\right)^{1/3}M_\odot.
\label{eq:msol_m_mh}
\end{equation}
We note that, from Eq.~\eqref{eq:rho_sol}, the cylindrical mass
profile within the projected radius $R$ is computed as
\begin{align}
  M_{\mathrm{cyl}}(<R) &=\int_0^R 2\pi
    R'dR'\int_{-\infty}^{\infty}\rho_{\mathrm{sol}}(\sqrt{R'{}^2+z^2})dz\nonumber\\
  &= M_{\mathrm{sol}}\left[1-\frac{1}{\{1+0.091(R/r_{\mathrm{c}})^2\}^{13/2}}\right],
\end{align}
which behaves as $M_{\mathrm{cyl}}(<R) \propto R^2$ at $R\ll
r_{\mathrm{c}}$ and $M_{\mathrm{cyl}}(<R) \rightarrow
M_{\mathrm{sol}}$ at $R\rightarrow \infty$.

\subsection{Low-mass perturber in JVAS B1938+666}
In the case of the $\sim 10^6M_\odot$ perturber in JVAS B1938+666, the
best-fitting model ('UD+PM' model) indicates the the mass of the
point-mass component of
$M_{\mathrm{PM}}=(4.25\pm 0.21)\times 10^5M_\odot$ and the mass of the
extended component of $M_{\mathrm{UD}}=(1.76 \pm 0.10)\times 10^6M_\odot$
\cite{2026NatAs..10..440V}. By interpreting them as an SMBH and a
soliton core, respectively, their mass ratio
$M_{\mathrm{SMBH}}/M_{\mathrm{sol}} \simeq 0.24$ is accurately known
from the observation, indicating that the effect of the SMBH on the
mass profile of the soliton core for this case can be evaluated 
quantitatively. The effect of an SMBH on the soliton core is
studied in detail in Ref.~\cite{2020MNRAS.492.5721D}, which we follow
to solve the Schr\"{o}dinger-Poisson equation with an SMBH potential
numerically, using the shooting method, to obtain the mass profile
(see also e.g., Ref.~\cite{2025arXiv250601282B}).  

\begin{figure}[tbp]
  \centering
\includegraphics[width=8.5cm]{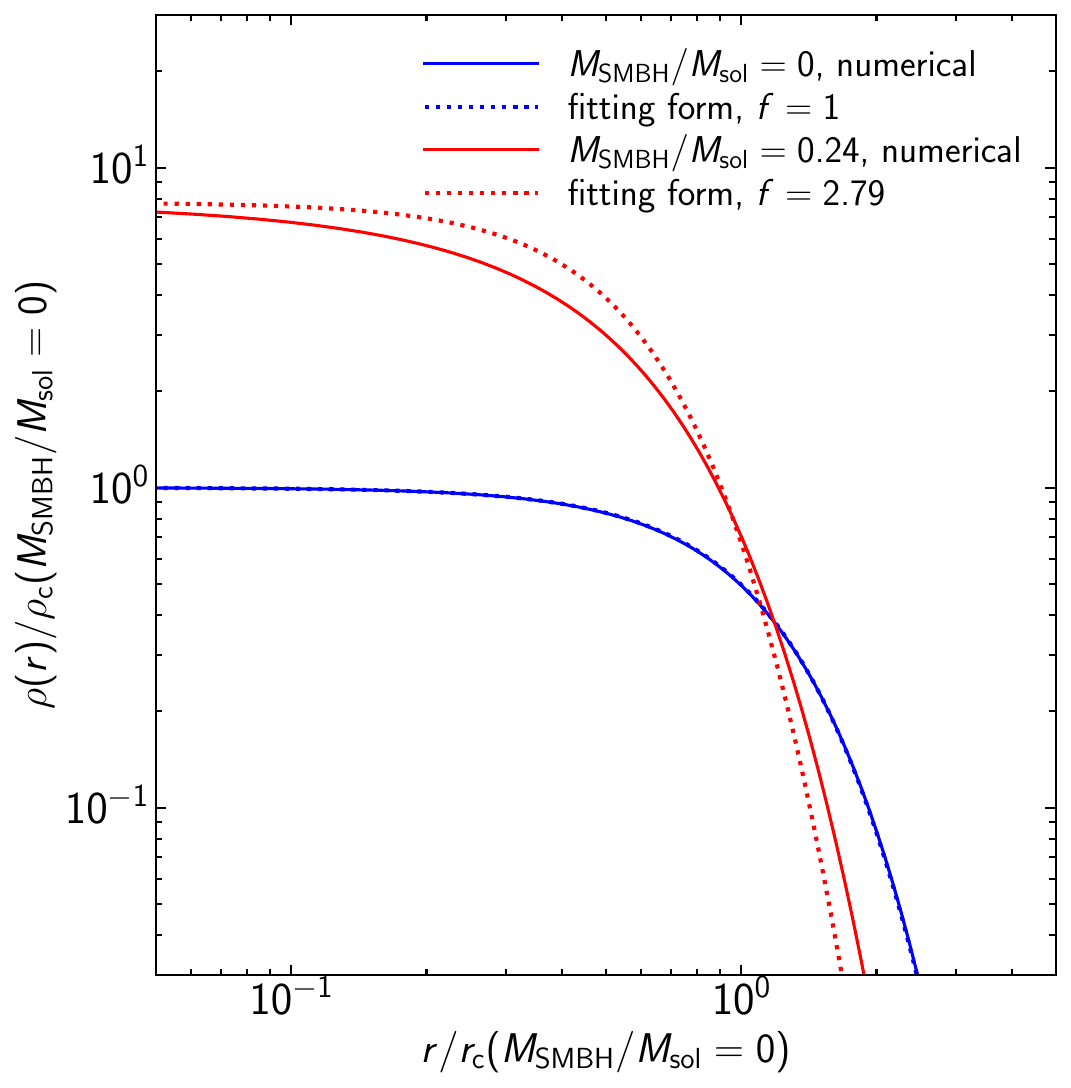}
\caption{\label{fig:solition} Mass profiles of the soliton core
  with ({\it red}) and without ({\it blue}) an SMBH with the ratio of
  the mass of the SMBH to that of the soliton core of
  $M_{\mathrm{SMBH}}/M_{\mathrm{sol}} =0.24$. Dotted lines indicate
  the fitting from given by Eq.~\eqref{eq:rho_sol} with the
  transformation of $\rho_{\mathrm{sol}}(r)\rightarrow
  f^2\rho_{\mathrm{sol}}(f^{2/3}r)$ with $f=1$ and $2.79$ for the case
  without and with an SMBH, respectively.}
\end{figure}

Fig.~\ref{fig:solition} compares mass profiles of the soliton core
with and without an SMBH, assuming that the total mass of the soliton
core is same for both cases. It is seen that the presence of an SMBH
enhances the gravitational attractive force to make the soliton core
more compact. Without an SMBH, we confirm that Eq.~\eqref{eq:rho_sol}
very well reproduces the numerical solution. Here we consider the
transformation
$\rho_{\mathrm{sol}}(r)\rightarrow f^2\rho_{\mathrm{sol}}(f^{2/3}r)$,
which conserves the total mass $M_{\mathrm{sol}}$, to check such a
transformed profile can reproduce the numerical solution for the case
of $M_{\mathrm{SMBH}}/M_{\mathrm{sol}} =0.24$. By choosing $f=2.79$
that corresponds to the value at $r\rightarrow 0$, we find that the
transformed version of Eq.~\eqref{eq:rho_sol} reproduces the numerical
solution reasonably well. This analysis suggests that the core radius
of the soliton core is reduced by a factor of $f^{-2/3}\simeq 0.50$
due to the presence of the SMBH with the mass ratio of
$M_{\mathrm{SMBH}}/M_{\mathrm{sol}} =0.24$.

\begin{figure}[tbp]
  \centering
\includegraphics[width=8.5cm]{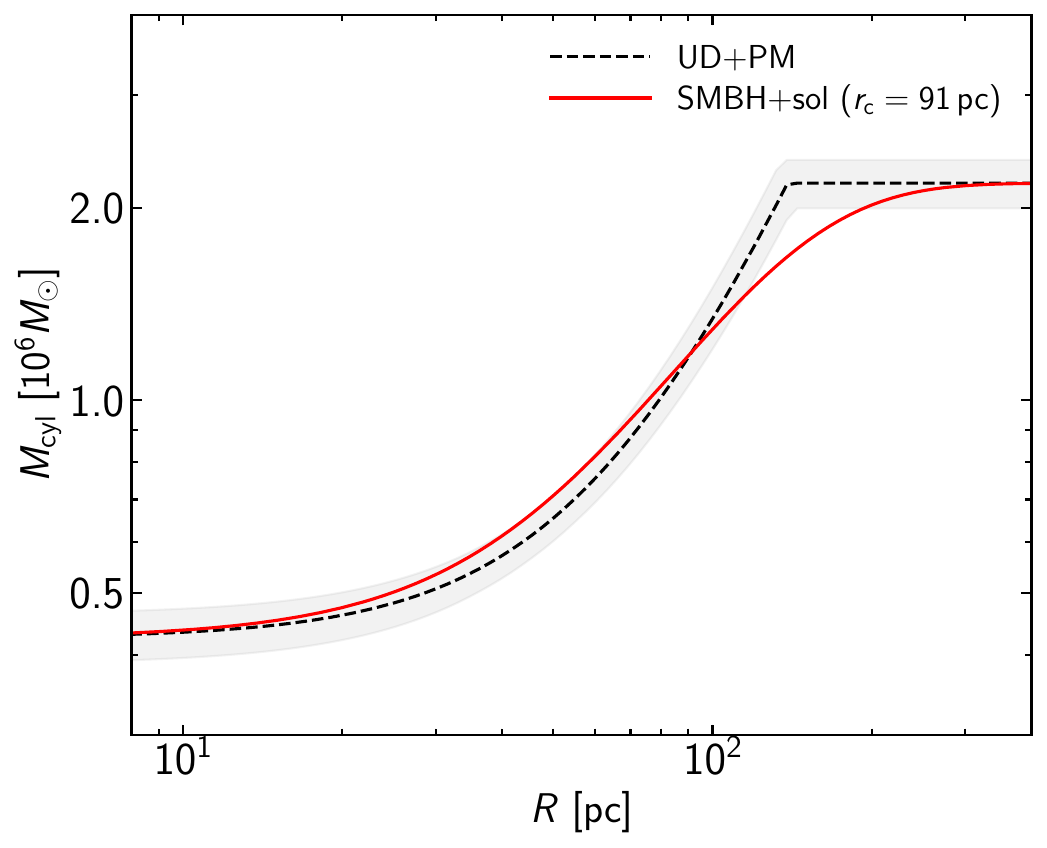}
\caption{\label{fig:mcyl} The cylindrical mass $M_{\mathrm{cyl}}(<R)$
  as a function of the projected radius $R$ for the UD+PM model,
  which corresponds to the best-fitting model in
  Ref.~\cite{2026NatAs..10..440V}, as well as $M_{\mathrm{cyl}}(<R)$
  for an SMBH plus a soliton core in the FDM model. The shaded
    region shows the $1\sigma$ error of $M_{\mathrm{cyl}}(<R)$ for the UD+PM
    model derived assuming that errors on the three model parameters
    given in Ref.~\cite{2026NatAs..10..440V} are not correlated.
  We choose $M_{\mathrm{SMBH}}=M_{\mathrm{PM}}$,
  $M_{\mathrm{sol}}=M_{\mathrm{UD}}$, and $r_{\mathrm{c}}=91$~pc.} 
\end{figure}

We then compare the cylindrical mass $M_{\mathrm{cyl}}(<R)$, which is
defined by the total mass of the lens within the projected radius $R$,
for our SMBH plus a soliton core model to check parameters that
reproduce the observed mass profile of the $\sim 10^6M_\odot$
perturber in JVAS B1938+666. We choose $M_{\mathrm{SMBH}}=M_{\mathrm{PM}}$ and 
$M_{\mathrm{sol}}=M_{\mathrm{UD}}$, assume Eq.~\eqref{eq:rho_sol} for
the density profile of the soliton core, and check what core radius
$r_{\mathrm{c}}$ can reproduce the observed mass profile.
We determine the soliton core radius so as to minimize $\chi^2$
  computed from the difference of $\log M_{\mathrm{cyl}}$ between the
  best-fitting UD+PM model and the SMBH plus a soliton core model to
  find the best-fitting soliton core radius of $r_{\mathrm{c}}\simeq
  91$~pc. The comparison in Fig.~\ref{fig:mcyl} suggests that the
  soliton core model can reproduce the sharp cut-off of the UD+PM
  model at outer radii reasonably well, at least much better compared
  with other physically-motivated profiles considered in
  Ref.~\cite{2026NatAs..10..440V}.
  We caution that conducting the forward strong lensing analysis
    as done in Ref.~\cite{2026NatAs..10..440V} assuming the soliton core
    profile will be crucial for assessing the validity of the soliton
    core model more quantitatively, which we leave for future work.
  In the following analysis, we adopt $r_{\mathrm{c,obs}}=91$~pc and
  explore scenarios that reproduce both $M_{\mathrm{UD}}$ and $r_{\mathrm{c,obs}}$.

\section{Possible scenarios}\label{sec:scenarios}

\subsection{FDM mass and halo mass}
\begin{figure}[tbp]
  \centering
\includegraphics[width=8.5cm]{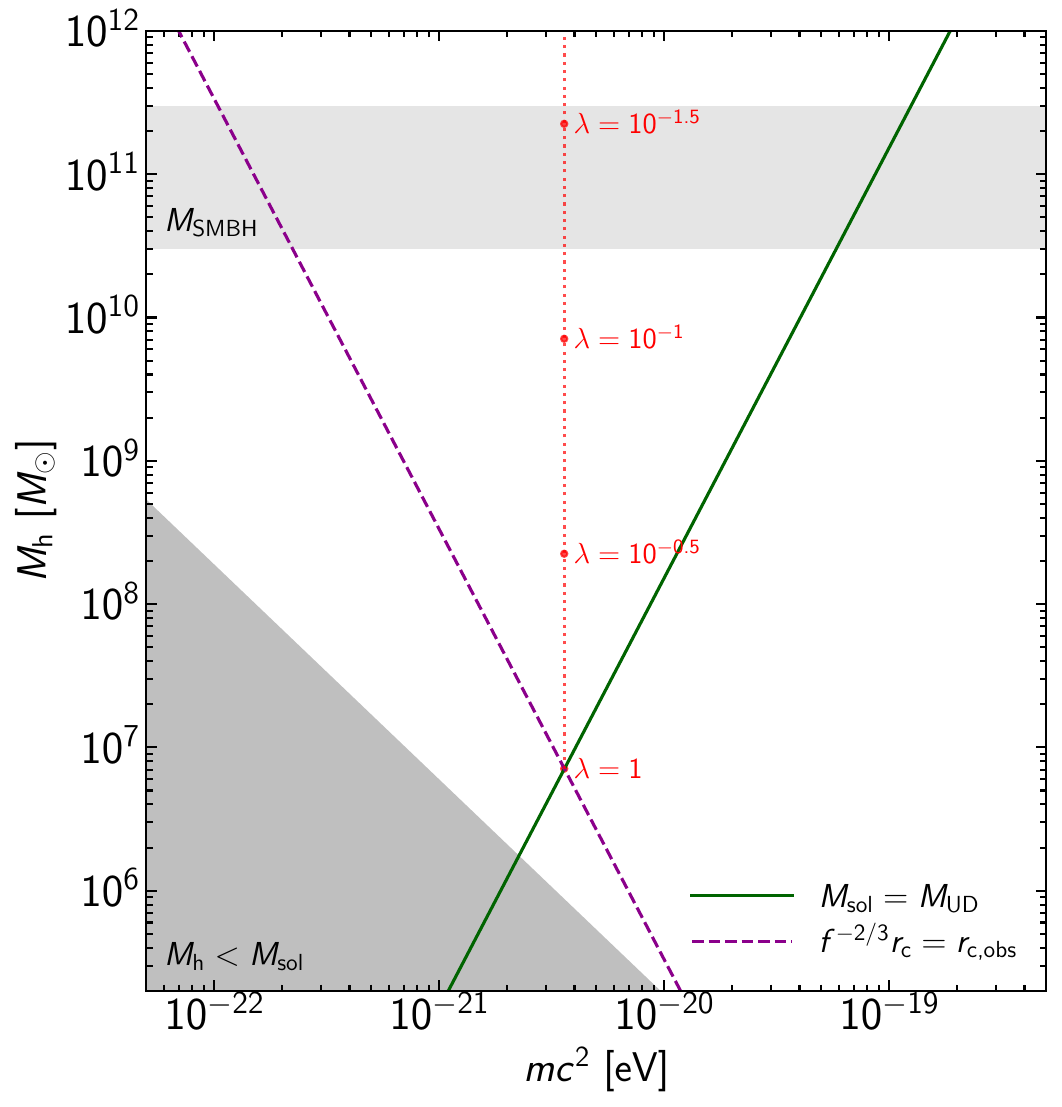}
\caption{\label{fig:mdm_mh} The FDM mass $mc^2$ and
  the halo mass $M_{\mathrm{h}}$ that explain the $\sim 10^6M_\odot$
  perturber in JVAS B1938+666. The dark shaded region is a forbidden
  region defined by $M_{\mathrm{h}}<M_{\mathrm{sol}}$. The filled
  point for $\lambda=1$ indicates the parameters that explain the
  observation in the absence of any tidal mass loss of the soliton
  core. The dotted line shows a track for different values of
  $\lambda$ that is defined by the ratio of the soliton core masses
  after and before the tidal mass loss. The light shaded region
  indicates a rough range of the halo mass that is consistent with
  the SMBH mass of $M_{\mathrm{SMBH}}=4.25\times 10^5M_\odot$ based on 
  the extrapolation of the SMBH mass--halo mass from higher masses.} 
\end{figure}

Here we discuss possible scenarios for the origin
of the $\sim 10^6M_\odot$ perturber in JVAS B1938+666.
The SMBH mass $M_{\mathrm{SMBH}}$, the total mass of the soliton
  core $M_{\mathrm{sol}}$, and the soliton core radius
  $r_{\mathrm{c,obs}}$ are determined by fitting the
  cylindrical mass profile of the best-fitting parametric UD+PM model
  in  Ref.~\cite{2026NatAs..10..440V}, as shown in
  Fig.~\ref{fig:mcyl}. More specifically, we assume
$M_{\mathrm{sol}}=M_{\mathrm{UD}}$ and
$f^{-2/3}r_{\mathrm{c}}=r_{\mathrm{c,obs}}$, where $f=2.79$ accounts
for the shrinking of the soliton core due to the SMBH for this
perturber. We then use Eqs.~\eqref{eq:msol_m_mh} and \eqref{eq:rc} 
to translate those soliton core mass and radius with the FDM mass
$mc^2$ and the halo mass $M_{\mathrm{h}}$, 
which we refer to the mass of a subhalo hosting the soliton core and
an SMBH and acting as the perturber in JVAS B1938+666, that
reproduce the observed mass profile, finding
$mc^2\simeq 3.6\times 10^{-21}$eV and
$M_{\mathrm{h}}\simeq 7.1\times 10^6M_\odot$ as shown in Fig.~\ref{fig:mdm_mh}. 

A potential issue is that the SMBH mass of
$M_{\mathrm{SMBH}}\simeq 4\times 10^5M_\odot$ may be too high
compared with the (sub)halo mass of $M_{\mathrm{h}}\simeq 7.1\times
10^6M_\odot$, where the halo mass is derived assuming the
  relation between the soliton core mass and the halo mass presented
  in Ref.~\cite{2014PhRvL.113z1302S}. For comparison, the SMBH mass--halo mass relation
constrained from observations (e.g., \cite{2023MNRAS.518.2123Z})
predicts that the halo mass corresponding to $M_{\mathrm{SMBH}}\simeq
4\times 10^5M_\odot$ is $M_{\mathrm{h}}\sim 10^{11}M_\odot$, which is
much higher than the halo mass needed to explain the soliton core.

\subsection{Tidal mass loss scenario}
A possible way to explain the high SMBH mass is to consider the
mass loss of the soliton core due to the tidal effect
\cite{2018PhRvD..97f3507D}. Once the halo that
hosts the soliton core and the SMBH enters the main halo that is
responsible for the primary strong lensing of JVAS B1938+666, the
tidal force acting on the soliton core affects the total mass and the
density profile of the soliton core. In previous studies (e.g.,
\cite{2018PhRvD..97f3507D}), it is shown that the density profile of
the soliton core after the mass loss due to the tidal evolution is
still approximately given by Eq.~\eqref{eq:rho_sol}. As a result, the
effect of the tidal evolution can be approximated by
the transformation $M_{\mathrm{sol}}\rightarrow \lambda M_{\mathrm{sol}}$
and $r_{\mathrm{c}}\rightarrow \lambda^{-1}M_{\mathrm{sol}}$ with
$\lambda<1$, which corresponds to the scaling transformation of
  the solution of the Schr\"{o}dinger-Poisson equation
  \cite{2014PhRvL.113z1302S}. Since the central density is changed as
$\rho_{\mathrm{sol}}(0)\rightarrow \lambda^4\rho_{\mathrm{sol}}(0)$,
the tidal mass loss makes the soliton core less dense and more
extended. By computing $M_{\mathrm{sol}}=M_{\mathrm{UD}}$ and
$f^{-2/3}r_{\mathrm{c}}=r_{\mathrm{c,obs}}$ after the transformation,
it is easily shown that the effect of the mass loss can increase the
halo mass $M_{\mathrm{h}}$ while the FDM mass $mc^2$ is
unchanged. Fig.~\ref{fig:mdm_mh} indicates that we may be able to
explain both the SMBH mass and the mass profile of the soliton core
simultaneously with $\lambda\sim 10^{-1.2}-10^{-1.5}$.

We can use the result in Ref.~\cite{2018PhRvD..97f3507D} to estimate
the efficiency of the mass loss of the soliton core. From
Eq.~\eqref{eq:rho_c}, we estimate the central density of the
soliton core without the gravitational effect of the SMBH to
$\rho_{\mathrm{sol}}(0)=0.025\lambda^{-4}M_\odot\mathrm{pc}^{-3}$. In
contrast, assuming the virial mass of the main lensing halo of JVAS B1938+666 to
$10^{13}M_\odot$ \cite{2014MNRAS.439.2494O} and adopting the
mass-concentration relation in Ref.~\cite{2019ApJ...871..168D}, the
halo density is
$\bar{\rho}=1.3\times 10^{-5}M_\odot\mathrm{pc}^{-3}$ at the virial radius and
$\bar{\rho}=0.042M_\odot\mathrm{pc}^{-3}$ at 1\% of the virial radius that
roughly corresponds to the Einstein radius. Therefore, for $\lambda<0.1$,
the ratio of the central density of the soliton core to the halo
density is $\mu=\rho_{\mathrm{sol}}(0)/\bar{\rho}> 6\times 10^6$ for
that range of radii. Since the tidal mass loss is not significant for 
$\mu\gtrsim 70$ \cite{2018PhRvD..97f3507D}, we conclude that the mass
loss scenario face difficulties in the lack of an efficient mass
loss. In addition, the abundance of massive subhalos is much smaller
compared with that of lower mass subhalos. Considering the discussion
on the abundance of the perturber in Ref.~\cite{2026NatAs..10..440V}
that the the observed abundance is in agreement with the abundance of
$10^6-10^7M_\odot$ subhalos in the standard cold dark matter (CDM)
scenario, the mass loss scenario also has difficulties in explaining
the observed abundance.  

\subsection{Heavy SMBH seed scenario}
We consider another scenario to explain the high SMBH mass, given
the fact that the current SMBH mass--halo mass relation is based on
observations of black holes with SMBH masses of $\gtrsim 10^7M_\odot$, and
hence the discussion above relies on the extrapolation of the  SMBH
mass--halo mass relation down to lower masses. Since the origin of
SMBHs is yet to be understood, it is possible that low mass halos host
SMBHs that are much more massive than predicted by the extrapolation
of the SMBH mass--halo mass relation. High SMBH masses in low mass
halos can be naturally explained in several scenarios for the origin
of SMBHs, including the direct collapse black hole formation
(e.g., \cite{2001ApJ...546..635O,2003ApJ...596...34B}) and
primordial black holes (e.g., \cite{2022ApJ...926..205C,2025ApJ...992..136Z}).
Such heavy SMBH seeds may also be suggested by recent discoveries of
little red dots (see e.g., \cite{2025arXiv251203130I}
for a review) {and highly over-massive black holes (e.g.,
  \cite{2024ApJ...960L...1N,2025JCAP...04..040Z}) at high redshifts.
The presence of an SMBH with $M_{\mathrm{SMBH}}\sim 10^{5}M_\odot$ in
a halo with $M_{\mathrm{h}}\sim 10^7M_\odot$ may be possible in such
scenarios. The observed
abundance of the perturber can also be naturally explained for those
scenarios given the low halo mass. 

\begin{figure}[tbp]
  \centering
\includegraphics[width=9.0cm]{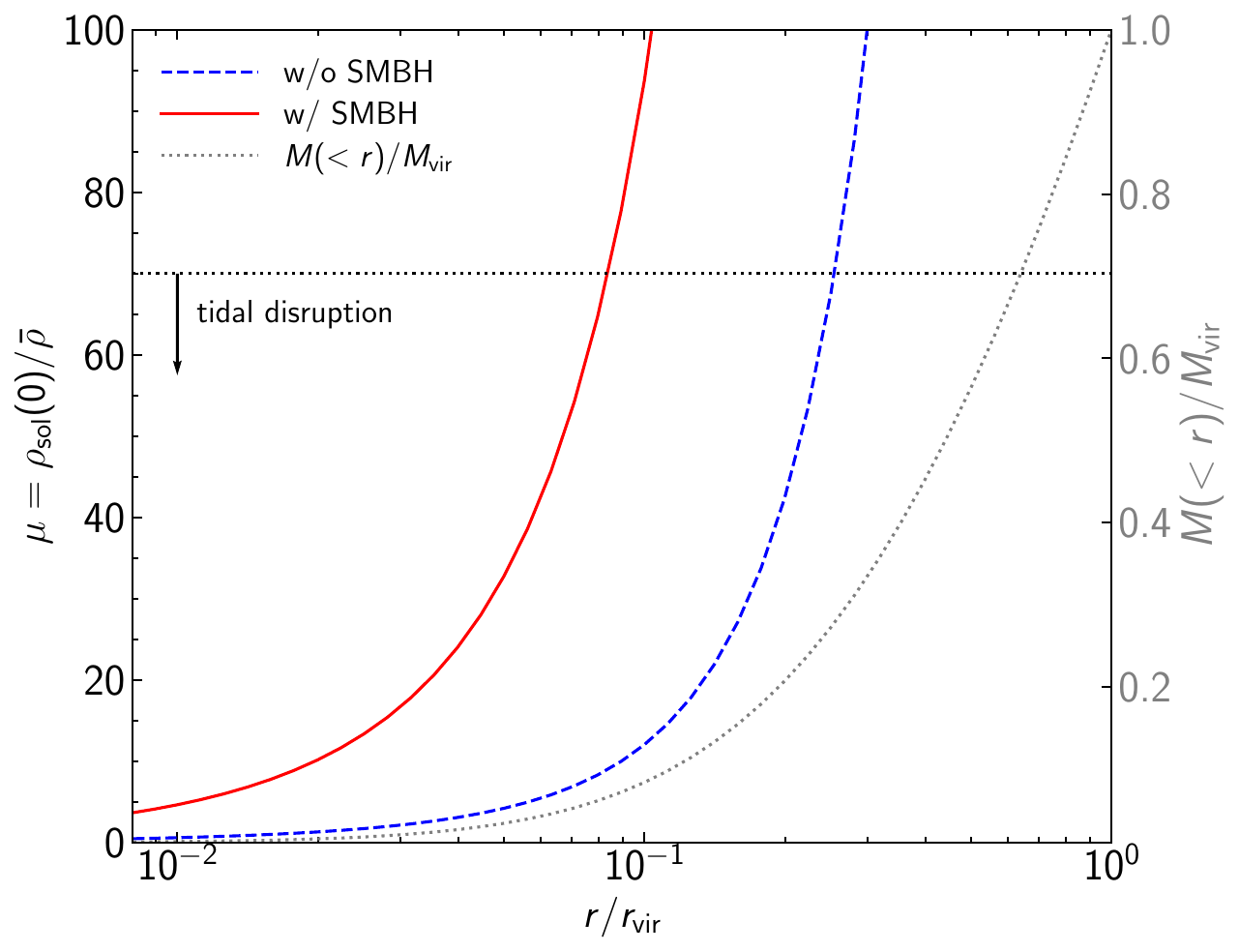}
\caption{\label{fig:mu} The density
  ratio $\mu=\rho_{\mathrm{sol}}(0)/\bar{\rho}$ between the central
  density of the soliton core with ({\it solid}) and without ({\it
    dashed}) the SMBH and the density of the main halo as a
  function of the distance $r/r_{\mathrm{vir}}$ of the soliton core
  from the center of the main halo. The horizontal dotted line
  indicates the rough condition for the tidal disruption, $\mu \lesssim 70$
  \cite{2018PhRvD..97f3507D}. The gray dotted line shows the enclosed
  mass fraction $M(<r)/M_{\mathrm{vir}}$ of the main halo as a
  function of $r/r_{\mathrm{vir}}$.} 
\end{figure}

Here we discuss the efficiency of the tidal mass loss for the soliton
core hosting the SMBH. As is clear from Fig.~\ref{fig:solition}, the
presence of the SMBH whose mass is comparable to the soliton core can
significantly enhances the central density of the soliton core, by a
factor of $f^2\simeq 7.8$, to make the mass loss much more 
inefficient. In Fig.~\ref{fig:mu}, we compare the density ratios
$\mu=\rho_{\mathrm{sol}}(0)/\bar{\rho}$ of soliton cores with
$M_{\mathrm{sol}}=M_{\mathrm{UD}}$ with and
without the SMBH as a function of their position within the main
halo. Assuming that the soliton cores are tidally disrupted for
$\mu \lesssim 70$, we find that the soliton cores survive at
$r/r_{\mathrm{vir}}\gtrsim 0.08$ and $\gtrsim 0.3$ with and without
the SMBH, respectively. While this analysis suggests that the soliton
core, even with the SMBH, is difficult to survive near the Einstein
radius of $r/r_{\mathrm{vir}}\sim 0.01$, we emphasize that strong
lensing perturbers can reside in the outskirt of the lensing halo as
long as their projected separation from the lens center is small.  
By checking the enclose mass fraction $M(<r)/M_{\mathrm{vir}}$ of the
main halo, this result indicates that roughly $\sim 5\%$ and $\sim
30\%$ of the soliton cores with and without the SMBH are tidally
disrupted, respectively. This analysis indicates that soliton cores
with central SMBHs are more likely to survive when they accrete onto
massive halos and they can be located closer to the center of the main
halo.

We note that here we implicitly assume that the effect of the tidal
mass loss is characterized by the density ratio $\mu$ alone even in the
presence of the SMBH. We discuss the validity of such assumption
in Appendix~\ref{sec:app}.

\section{Summary}\label{sec:summary}

Ref.~\cite{2026NatAs..10..440V} argues that the mass profile of the
$\sim 10^6M_\odot$ perturber in JVAS B1938+666 does not resemble that
of any known astronomical object. We advocate that the mass profile is
naturally explained by considering a soliton core in the FDM model
hosting an SMBH. We have found that, given the observed mass ratio of
the point mass and extended components, the gravitational force of the
SMBH affects the mass profile of the soliton core such that its core
radius is reduced by a factor of $\sim 2$ and its central density is
enhanced by a factor of $\sim 8$. The observed mass profile of the
soliton core is explained by the FDM mass of
$mc^2\simeq 3.6\times 10^{-21}$eV and the halo mass of
$M_{\mathrm{h}}\simeq 7.1\times 10^6M_\odot$. This halo mass is
  found to be several orders of magnitudes smaller than that expected
  from the standard SMBH mass--halo mass relation, $M_{\mathrm{h}}\sim
  10^{11}M_\odot$, for the mass of the SMBH of
  $M_{\mathrm{SMBH}}=4.25\times 10^5M_\odot$, even though the SMBH
  mass--halo mass relation is currently not constrained tightly at
  such low  mass. We argue that the small halo mass in our model may 
be reconciled for some scenarios that predict heavy SMBH seeds, such
as the direct collapse black hole formation and primordial black holes.
Our model is physically motivated and approximately explains the
  sharp cut-off and the constant density of the unphysical 'UD' model
  considered in Ref.~\cite{2026NatAs..10..440V}. 

We note that are several caveats in this work. First, there are
several observations that tightly constrain the FDM model such that
the FDM model cannot explain all the dark matter for a wide range of
the FDM mass (e.g., \cite{2025arXiv250700705E} for a
review). Therefore the scenario proposed in this paper may not work as
it is, and some modifications such as a compound dark matter model
consisting of the FDM and other form of dark matter, considering the
two-field (or 
multi-field) FDM model, or including the self-interaction of the FDM may be
needed \cite{2025arXiv250700705E}. For instance, simulations
  suggest that soliton cores still exist in the compound dark matter model
  even if the FDM accounts for 10\% or less of the total dark matter
  density \cite{2025MNRAS.537..252D}, although in such a scenario the
  corresponding halo mass would become higher considering the smaller
  soliton core mass for a given halo mass. Second, discussions in this paper
rely on the relation between the soliton core mass and the halo mass in
Ref.~\cite{2014PhRvL.113z1302S}, which is subject to various 
uncertainties and diversities (e.g.,
\cite{2021MNRAS.506.2603M,2021MNRAS.501.1539N,2022MNRAS.511..943C}).
More specifically, while the soliton core masses and the core
  radius are directly constrained from the observation, the inferred
  FDM mass and the halo mass rely on the assumption on the relation
  between the soliton core mass and the halo mass. 
Third, other systematic effects such as uncertainties of the estimation of
the tidal mass loss \cite{2025MNRAS.540.2653C} and barynoic effects
should also affect our quantitative result.

Despite these caveats, we expect that our argument that a soliton core
in the FDM model serves as a more efficient perturber in a strong lens
system when it hosts an SMBH at the center generically holds. Since
the SMBH enhances the central density of the soliton core, it is less
easily disrupted by the tidal effect and its lensing effect is more
significant. Such dens low-mass perturbers will affect strong lensing
observables in several ways, including flux ratio anomalies (e.g.,
\cite{2020MNRAS.491.6077G}) and positional anomalies (e.g.,
  \cite{2023NatAs...7..736A}). We leave the exploration along this
  line to future work.

\acknowledgments

We thank the anonymous referee for useful comments and suggestions.
This work was supported by JSPS KAKENHI Grant Numbers JP25H00662, JP25H00672.

\appendix
\section{Effect of the tidal mass loss of an SMBH}
\label{sec:app}
  
The presence of an SMBH in the soliton core can affect the tidal mass
loss mainly in two ways, one is the enhancement of the central density
of the soliton core and the other is the contribution of the
gravitational potential of the SMBH on the total potential, both of
which are expected to make the tidal mass loss less efficient. In
Sec.~\ref{sec:scenarios}, we assume that the latter effect is
negligible. Here we check the validity of such approximation using the
WKB approximation. 

The tidal mass loss of a soliton core can be studied analytically
with the Schr\"{o}dinger-Poisson equation including the tidal
potential \cite{2017PhRvD..95d3541H,2018PhRvD..97f3507D}. In the
presence of an SMBH with the mass $M_{\mathrm{SMBH}}$, it is given by 
\begin{equation}
i\hbar\frac{\partial \psi}{\partial t}=\left[-\frac{\hbar^2}{2m}\nabla^2+m\left(\Phi-\frac{GM_{\mathrm{SMBH}}}{r}-\frac{3}{2}\omega^2r^2\right)\right]\psi,
\end{equation}
\begin{equation}
\nabla^2\Phi=4\pi G |\psi|^2,
\end{equation}
where $\Phi$ denotes a gravitational potential due to the self-gravity
of the soliton core and $\omega$ is the angular speed of the soliton
core moving within a main halo. We note that $\omega$ is connected
with the halo density at the position of the soliton core and hence
the ratio of the central density of the soliton core to the halo
density $\mu$. The effective potential of this system is given by
\begin{equation}
V_{\mathrm{eff}}(r)=m\Phi-m\frac{GM_{\mathrm{SMBH}}}{r}-\frac{3}{2}m\omega^2r^2.
\end{equation}
Even when the energy eigenvalue $E_0$ satisfies the classical
stability condition $\max(V_{\mathrm{eff}}) > E_0$, FDM particles can
penetrate the potential barrier via the tunneling effect. We evaluate
this decay lifetime using the WKB approximation. The Gamow factor
$\gamma$ in the classically forbidden barrier region is given by
\begin{equation}
\gamma = \int_{r_{\mathrm{in}}}^{r_{\mathrm{out}}} \frac{\sqrt{2(V_{\mathrm{eff}}(r)-E_0)}}{\hbar}\, dr,
\end{equation}
where $r_{\mathrm{in}}$ and $r_{\mathrm{out}}$ are the classical
turning points satisfying $V_{\mathrm{eff}}(r)=E_0$. Since the
probability $T$ of tunneling through this barrier is approximated
by $T \simeq \exp(-2\gamma)$, the decay lifetime $\tau$ of the soliton
core is proportional to the inverse of $T$ as
$\tau \propto \exp(2\gamma)$. Let $\gamma_0$ and $\tau_0$ be the Gamow
factor and the decay lifetime calculated in the absence of the SMBH
potential but with the enhanced density profile (and hence $\Phi$) due
to the SMBH. The enhancement of the decay lifetime due to the presence of
the SMBH potential is expressed as 
\begin{equation}
\log_{10} \left( \frac{\tau}{\tau_0} \right) = \frac{2(\gamma - \gamma_0)}{\ln 10}.
\end{equation}

\begin{figure}[tbp]
  \centering
\includegraphics[width=8.0cm]{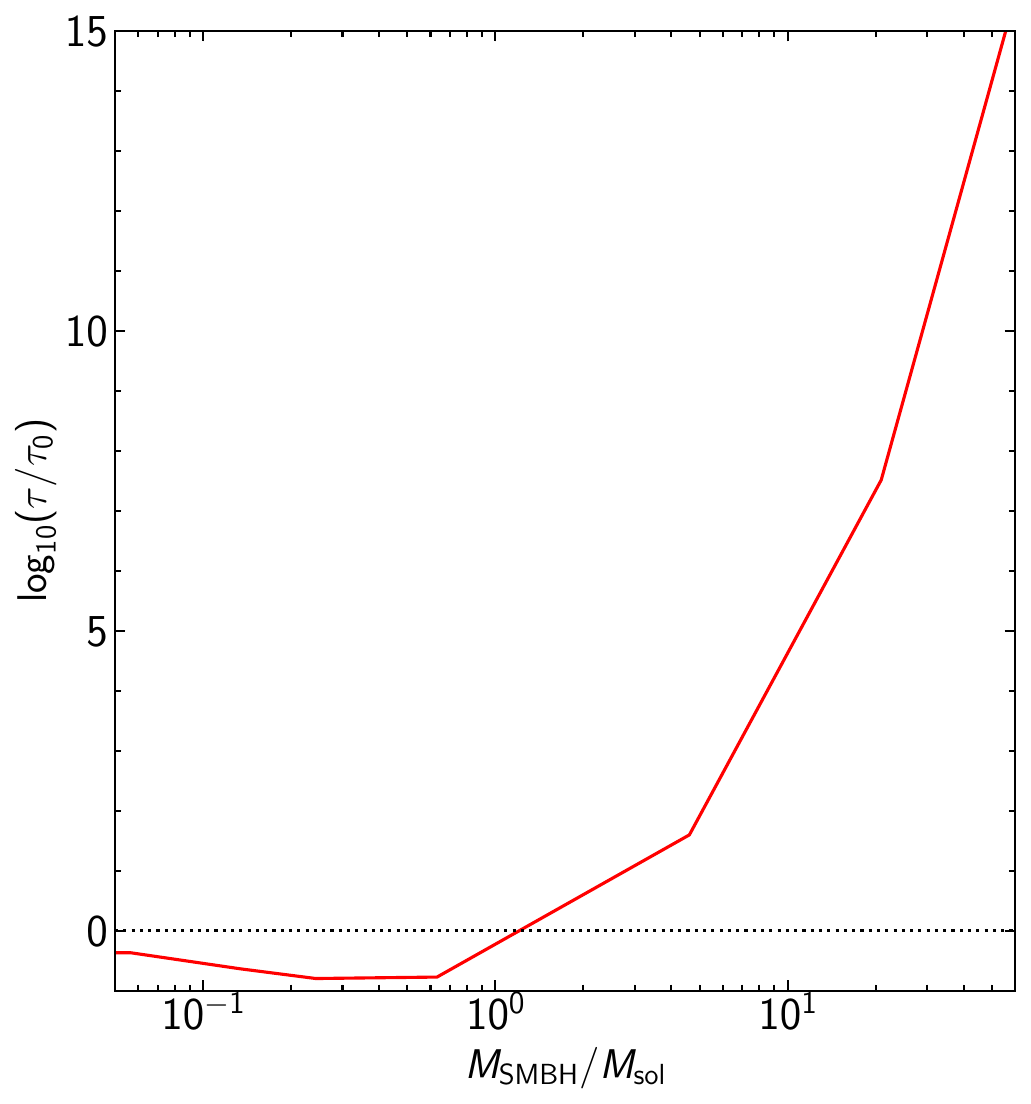}
\caption{\label{fig:tau} Dependence of the decay lifetime enhancement
  $\log_{10}(\tau/\tau_0)$ of the soliton core on the mass ratio
  $M_{\mathrm{SMBH}}/M_{\mathrm{sol}}$, assuming the ratio of the
    central density of the soliton core to the halo density of $\mu=70$. } 
\end{figure}

In Fig.~\ref{fig:tau}, we show the enhancement of the decay lifetime due to the
SMBH potential. We find that $\log_{10}(\tau/\tau_0)$ is close to $0$
for $M_{\mathrm{SMBH}}/M_{\mathrm{sol}}\lesssim 5$, indicating that
the effect of the SMBH potential is not significant and that the
effect of an SMBH on the tidal mass loss mainly comes from the
enhancement of the central density of the soliton core due to the
SMBH in this range of the mass ratio, considering the very steep
dependence of the tidal mass loss rate on $\mu$
\cite{2017PhRvD..95d3541H,2018PhRvD..97f3507D}. Since the mass ratio
of our interest in this study is $M_{\mathrm{SMBH}}/M_{\mathrm{sol}}=0.24$,
we conclude that our approximation to ignore the effect of the SMBH
potential is valid. On the other hand, when
$M_{\mathrm{SMBH}}/M_{\mathrm{sol}}\gtrsim 5$, the soliton core is
strongly bound by the SMBH, preventing it from undergoing classical
disruption even in a strong tidal field. In such a situation, we need
to take proper account of the SMBH potential in calculating the tidal
mass loss rate.

\bibliographystyle{JHEP}
\bibliography{ref}

\end{document}